\documentclass[%
reprint,
amsmath,
amssymb,
aps,
prl,
floatfix,
superscriptaddress,
longbibliography,
]{revtex4-2}

\usepackage{makecell}
\usepackage{graphicx}
\usepackage[utf8]{inputenc}
\usepackage{dcolumn}
\usepackage{bm}
\usepackage[colorinlistoftodos]{todonotes}
\presetkeys{todonotes}{inline}{}
\usepackage{chemformula}
\usepackage{xspace}
\usepackage{mathtools}
\usepackage{siunitx}
\usepackage{notes2bib} 
\usepackage[capitalise]{cleveref}
\usepackage{xcolor}

\bibnotesetup{
note-name = ,
use-sort-key = false
}

\definecolor{newtext}{RGB}{255, 0, 0}
\definecolor{AMK}{RGB}{20, 20, 220}

\crefname{subsection}{subsection}{subsections}

\newcommand{\cm}{cm$^{-1}$\xspace}

\newcommand{\TN}{$T_N$\xspace}
\newcommand{\Tcr}{$T^*$\xspace}

\newcommand{\CoF}{\ch{CoF2}\xspace}
\newcommand{\MnF}{\ch{MnF2}\xspace}

\newcommand{\FeF}{\ch{FeF2}\xspace}

\newcommand{\Ag}{$A_{1g}$\xspace}
\newcommand{\Bone}{$B_{1g}$\xspace}

\newcommand{\Eg}{$E_{g}$\xspace}

\begin{document}

\title{Giant intrinsic nonlinear phonon - magnon coupling in the antiferromagnet \CoF}

\author{M. A. Prosnikov}
    \email{yotungh@gmail.com}
    \affiliation{Ioffe Institute, 194021 St.-Petersburg, Russia}
    \affiliation{High Field Magnet Laboratory (HFML--EMFL), Radboud University, Toernooiveld 7, 6525 ED Nijmegen, The Netherlands}
    \affiliation{Radboud University, Institute for Molecules and Materials, Heyendaalseweg 135, 6525 AJ Nijmegen, The Netherlands}
\author{M. Bal}
     \affiliation{High Field Magnet Laboratory (HFML--EMFL), Radboud University, Toernooiveld 7, 6525 ED Nijmegen, The Netherlands}
     \affiliation{Radboud University, Institute for Molecules and Materials, Heyendaalseweg 135, 6525 AJ Nijmegen, The Netherlands}
\author{R. V. Pisarev}
     \affiliation{Ioffe Institute, 194021 St.-Petersburg, Russia}
\author{P. C. M. Christianen}
     \affiliation{High Field Magnet Laboratory (HFML--EMFL), Radboud University, Toernooiveld 7, 6525 ED Nijmegen, The Netherlands}
     \affiliation{Radboud University, Institute for Molecules and Materials, Heyendaalseweg 135, 6525 AJ Nijmegen, The Netherlands}
\author{A. M. Kalashnikova}
     \affiliation{Ioffe Institute, 194021 St.-Petersburg, Russia}
\date{\today}

\begin{abstract}
The observation of strongly coupled lattice and spin dynamics in altermagnet \CoF is reported.
On the background of the expected spin-phonon interaction leading to renormalization of all phonons at the N\'eel temperature an additional strong coupling between one-magnon excitation and the lowest frequency Raman-active phonon of \Bone symmetry was observed and manifested an anomaly in its energy, full width, and intensity at temperature \Tcr$=23~K$ precisely where the frequency of the phonon becomes exactly twice the frequency of the magnon.
We assigned this effect to unique magnon-phonon coupling in the form of a two-magnons-one-phonon interaction.
The consistent experimental data clearly demonstrate that there is an intrinsic coupling that does not require coherent excitation. 
\end{abstract}


\maketitle

\section{\label{sec:intro}Introduction}
Coupling of the spin to lattice dynamics is one of the dominant mechanisms responsible for and limiting the lifetimes of coherent magnetic excitations -- magnons~\cite{bozhko_magnonphonon_interactions_2020,valdesaguilar_colossal_magnonphonon_2007,streib_magnonphonon_interactions_2019,kim_bonddependent_anisotropy_2023}.
Therefore, revealing mechanisms responsible for such a coupling and suggesting pathways for suppressing it is of high interest for developing magnonic devices of the future~\cite{bozhko_magnonphonon_interactions_2020,pirro_advances_coherent_2021,juraschek_magnetic_control_2021}.
On the other hand, terahertz lattice dynamics, which readily couples to electromagnetic waves, enables driving magnetic excitations at such frequencies, which has been recently recognized as a promising approach to optical excitation of coherent THz spin modes~\cite{nova_effective_magnetic_2017}, forced oscillations~\cite{formisano_laserinduced_thz_2022}, manipulation of the exchange interactions~\cite{afanasiev_ultrafast_control_2021, fechner_magnetophononics_ultrafast_2018}, as well as other types of ultrafast spin dynamics \cite{disa_polarizing_antiferromagnet_2020,cartella_parametric_amplification_2018}.

In most of the cases described, either magnon coupling to low-frequency acoustic modes~\cite{dabrowski_coherent_transfer_2020} in the form of parametric resonance or effects achieved via direct energy overlap between excitations are considered~\cite{valdesaguilar_colossal_magnonphonon_2007}, which usually makes these effects hard to study due to the inability to disentangle different contributions.

Recently rutiles were proposes as model altermagnetic systems with \CoF being one of them~\cite{guo_spinsplit_collinear_2023,adamantopoulos_spin_orbital_2024,bhowal_ferroically_ordered_2024}.  Morover, several pathways to exploit phonon-magnon coupling were realized with a rich spectrum of lattice and magnetic excitations.
First, in \cite{disa_polarizing_antiferromagnet_2020} and \cite{formisano_laserinduced_thz_2022}, coherent phonons of $B_{2g}$ and \Eg were employed to induce quasi-stationary and THz nonequlibrium spin alignments, respectively.
On the other hand, in \cite{mashkovich_terahertz_light_2021}, magnons coherently driven by a strong THz pulse enabled the generation of the \Bone phonons via two-magnon conversion.

However, all these experiments follow a scenario where one of the excitations, either the phonon or the magnon, is resonantly driven and has a pronounced amplitude, while the other is seen only when the energy and symmetry conditions are satisfied for its coupling to the driven mode.
Thus, such studies essentially do not allow investigation of both modes under the same conditions when the coupling between them is either present or absent and thus provide only fragmented information on the impact of phonon-magnon coupling on the properties of these modes.

Moreover, the discussed works \cite{disa_polarizing_antiferromagnet_2020,formisano_laserinduced_thz_2022,mashkovich_terahertz_light_2021} focus on the effects involving coherent and high-amplitude excitations created by ultrashort pulses in the optical or THz range, and the obtained results cannot be directly applied to systems close to equilibrium.
Only for spin dynamics driven in \CoF because of coupling to the coherently driven \Eg phonon mode \cite{formisano_laserinduced_thz_2022}, an incoherent counterpart has been reported \cite{allen_magnetic_excitations_1971}.

In this Letter, we address the following question: could the strong magnon-phonon coupling between the \Bone phonon and the magnon in \CoF be fully intrinsic effect?
In order to do so, the inherently non-resonant and non-coherent spontaneous Raman scattering technique was used to rigorously explore the symmetry, frequencies, and lifetimes of the \Bone phonon and magnon modes in \CoF.
Moreover, the remarkable spectral resolution and direct access to modes of different symmetry via the polarization combination of incident and scattered light of the chosen technique are ideal for unraveling convoluted spin/lattice dynamics of \CoF. 
By tuning the temperature, we examine fine changes in the characteristics of the lattice and magnetic modes and reveal their evolution as the condition for nonlinear resonance between \Bone and a pair of magnons is reached.
We show that the coupling between two magnons and one \Bone phonon manifests itself in the whole range of temperatures below~\TN$=38$~K, as shown in~\cref{fig:magnetic_temp}.
Independent observation of the phonon and magnon modes in spontaneous Raman scattering far and close to the nonlinear resonance condition, enabled us to demonstrate that the coupling has a pronounced mutual effect on both lattice and magnetic dynamics.
This allowed us to demonstrate that the phonon - two-magnon coupling in \CoF possesses an intrinsic nature and does not require selective and coherent excitation of any of the modes.

\section{\label{sec:results}Results and discussion}

\begin{figure}
    \centering
        \includegraphics[width=1.0\columnwidth]{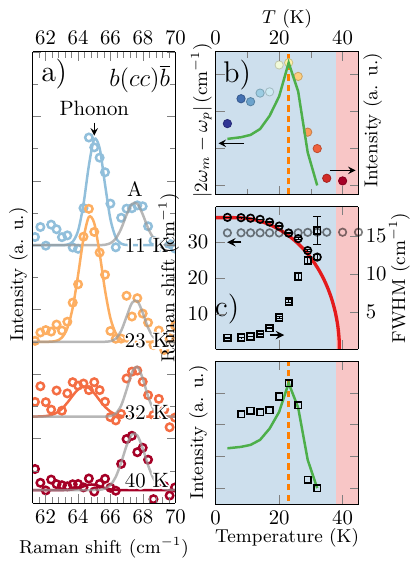}
	\caption{
 	\label{fig:magnetic_temp}
        a) Temperature dependence Raman spectra line measured in the ``forbidden'' geometry $b(cc)\overline{b}$ at several temperatures demonstrating emergence of the \Bone mode below \TN.
        b) Temperature dependence of the integral intensity  of the \Bone mode demonstrating non-monotonous behaviour with maxima at \Tcr.
        Green line shows the evolution of the detuning between \Bone phonon frequency and doubled magnon frequency, $|2\omega_\mathrm{m}-\omega_p|$.
        c) Raman shift (black circles) and FWHM (squares), in comparison with half of the frequency of the \Bone phonon mode (gray circles).
        Red line depicts the scaled Brillouin $\langle S \rangle$ dependence.
        Dashed orange vertical line depicts \Tcr, where coupling occurs.
        d) integral intensity of the one magnon-mode in comparison with the detuning.
	}
\end{figure}

\begin{figure}
    \centering
	\includegraphics[width=1.0\columnwidth]{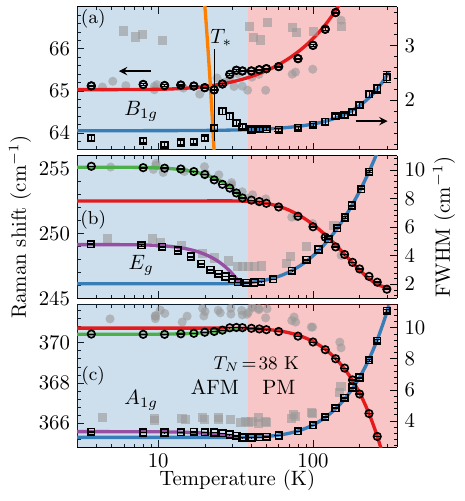}
	\caption{
	\label{fig:spin_phonon} 
        (b-d) Temperature dependence of the frequency (circles, left-hand axis) and FWHM, (squares, right-hand axis) of the three lowest-lying phonon modes (b) \Bone, (c) \Eg, and (d) \Ag.
        Blue and red regions correspond to antiferromagnetic and paramagnetic phases, respectively.
        Red and blue solid lines correspond to temperature evolution of $\omega_{ph}$ and FWHM due to anharmonic four-phonon processes \bibnote{see~\cref{eq:freq,eq:FWHM} in SM}, while green and violet ones also account for spin-phonon contributions.
        Gray circles and squares depict data from~\cite{cottam_Spinphonon_Interaction_2019}.
        (a) shows a crossing of the \Bone phonon and doubled frequency of the one-magnon excitation (solid orange line) at \Tcr where anomalies occur (see text for details).
    }
\end{figure}

\begin{figure}
    \centering
	\includegraphics[width=1.0\columnwidth]{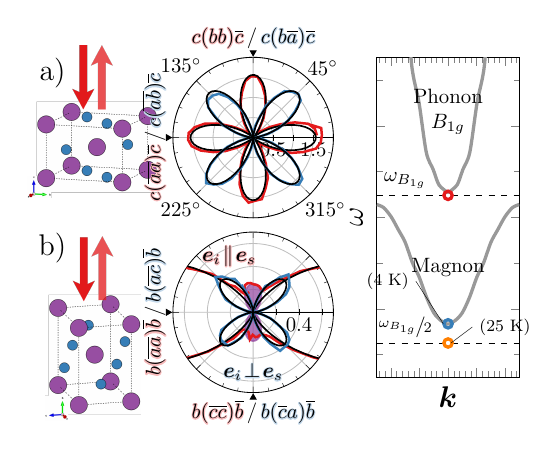}
	\caption{
        \label{fig:polar_maps}
        a-b) Crystal structure of the \CoF with incident and scattered light shown as arrows.    
        c-d) Azimuthal dependence of the integral intensity of the \Bone phonon line in the Raman spectrum as measured at $T=4.2$~K in parallel (red) and crossed (blue) configurations.
        (b) Zoomed-in azimuthal dependence at $\mathbf{k}\parallel b$ demonstrating appearance of the phonon line in the forbidden configuration (shaded area) due to a dynamical symmetry breaking caused by phonon-magnon coupling.
        Black lines show fits using~\cref{eq:Raman intensity}.
        c) The dispersion curves of the magnon (at 4 K) and phonon (at 80 K) measured by inelastic neutron scattering adopted from \cite{martel_experimental_studies_1968} shown as gray lines.
        Our observations in the Brillouin zone center are shown as colored circles.
        }
\end{figure}

To reveal the coupled phonon-magnon dynamics, we studied high-quality single crystals of \CoF with the use of high-resolution Raman spectroscopy (resolution of about 0.3~\cm), which allowed us to directly and simultaneously monitor even subtle changes in the spectral parameters of the phonon and magnon modes, while azimuthal resolution provides direct information about their symmetry.

First of all the analysis of the symmetry of all Raman-active phonon and magnon modes in the spectral range of 20--400~\cm~ was done (see~\cref{fig:appendix_temperature_map} in the supplementary for the complete spectra).
While the selection rules and most of the parameters have already been established for \CoF~\cite{meloche_twomagnon_inelastic_2007,barreda-argueso_Pressureinduced_Phasetransition_2013}, here, for the first time, we investigate the full azimuthal dependencies of the integral intensities (area under the peak) of the corresponding lines in the Raman spectra in both the paramagnetic and antiferromagnetic phases, using the Raman tensor formalism~\cite{kranert_raman_tensor_2016}.
In the paramagnetic phase, our results for the frequencies and symmetry of the observed \Bone, \Ag, and \Eg  phonon modes are in full agreement with previously reported studies~\cite{macfarlane_light_scattering_1970, barreda-argueso_crystalfield_theory_2016}, and goes beyond that by providing a quantitative insight in the values of the particular Raman tensor elements, as demonstrated in~\cref{eq:element_values}.

However, below the  N\'eel temperature \TN we reveal the anomaly in the azimuthal dependence of the integral intensity of the phonon with \Bone symmetry.
Figure~\ref{fig:polar_maps}(a,b) shows the azimuthal dependence of the Raman line corresponding to the \Bone phonon mode obtained at $T=4.2$~K in four configurations.
According to its Raman tensor~\cref{eq:raman_tensors:B1g}, these dependencies are anisotropic and should follow the specific pattern shown in~\cref{fig:polar_maps} as black curves.
Indeed, for almost all configurations, the experiments are consistent with the theoretical predictions.
However, there is a notable nonzero intensity in the configuration $b(cc)\overline{b}$, which is one of the ``forbidden'' polarizations, according to the Raman tensor of the \Bone phonon mode \cref{eq:raman_tensors:B1g}.
This intensity is about 15\% of that in the allowed polarization $b(aa)\overline{b}$, as shown in~\cref{fig:polar_maps}~(b), which is well above the detection limit and surely exceeds ``leakages'' eventually found for other modes.

To get further insight, a temperature-dependent study was conducted, which revealed that the intensity of the \Bone mode in the forbidden polarization is strongly temperature dependent (see \cref{fig:polar_maps}~(c)).
It emerges below \TN, rapidly grows, and reaches a maximum at $T^*=23$~K, followed by a partial decrease and saturation at $T<15$~K, as shown in \cref{fig:polar_maps}~(d).
However, the frequency $\omega_\mathrm{B1g}=65$~\cm of this mode is simultaneously well above the one-magnon mode frequency, $\omega_\mathrm{m}=38$~\cm and well below the two-magnon band at 115~\cm.
Thus, there should be no resonant intensity transfer between the \Bone phonon mode and any of these magnetic excitations via direct coupling.

Previously, an activation of the even \Eg phonon at 256~\cm was reported in the far-infrared absorption spectra, where it should be forbidden by symmetry, but nevertheless occurs due to single-ion-orbit--lattice interaction~\cite{allen_spinopticalphonon_interaction_1968,mills_exciton_opticalphonon_1970}.
However, no such effect is expected for the \Bone mode, as it is far away from the frequency overlap and substantially narrower.
On the other hand, the emergence of the \Bone phonon mode intensity in the ``forbidden'' Raman geometry below \TN with non-monothonous temperature dependence closely resembles the behavior of the \Bone mode when it is driven by a strong THz pulse via the coherent two-magnon process reported in~\cite{mashkovich_terahertz_light_2021}.
Thus, our results suggest that nonlinear coupling between the \Bone phonon and magnon modes has an intrinsic nature and is feasible without a strong external driving force as well.

Figure~\cref{fig:spin_phonon}(b) shows the detailed temperature dependence of the frequencies and full widths at a half-maximum (FWHM) of the lines in the Raman spectra corresponding to the \Bone phonon mode.
As a reference, \cref{fig:spin_phonon}(c,d) shows the same parameters for the modes \Eg and \Ag, for which no anomalies in the azimuthal dependencies below \TN were found.
Note that, owing to the high spectral resolution of our experiments, the values of FWHM for the \Bone mode are found to be four times lower in comparison to those reported in one of the latest published works on Raman scattering in \CoF~\cite{cottam_Spinphonon_Interaction_2019} and shown in~\cref{fig:spin_phonon} by gray symbols.
Evidently, the enhanced spectral resolution of our data and superior signal-to-noise ratio enable us to resolve and analyze important features of these temperature dependencies, especially for the \Bone phonon mode, as discussed below.

First, we discuss the paramagnetic (PM) phase, where the leading effect defining the temperature dependencies of the frequency and FWHM of the phonon modes is the lattice anharmonicity approached by the theory described in~\cite{cottam_Spinphonon_Interaction_2019,balkanski_infrared_lattice_1966}.
As temperature decreases, the anharmonism is ``frozen out'', leading to a monotonous increase in the phonon frequency and their lifetimes, as seen for the \Eg and \Ag phonon modes (\cref{fig:spin_phonon}(c,d)).
However, as it was previously observed for most of the transition metal difluorides, the \Bone mode behaves in a counter-intuitive way, namely that its frequency drops with decreasing temperature.
Similar behavior was also observed in high-pressure experiments~\cite{barreda-argueso_Pressureinduced_Phasetransition_2013}, and it was concluded that a specific normal mode corresponding to the rotation of the fluorine ions forming octahedron around the metal cation led to a negative Gr\"uneisen parameter and pressure coefficient~\cite{barreda-argueso_Pressureinduced_Phasetransition_2013,chatterji_magnetoelastic_effect_2010a}.

Further, at lower temperatures below \TN, the spin-phonon effect starts to contribute, caused by the appearance of long-range AFM order, which modifies the force constants of the lattice excitations~\cite{cottam_Spinphonon_Interaction_2019}, and leads to the renormalization of the phonon frequency $\delta\omega_\mathrm{ph}\propto\lambda \langle S S \rangle$, with $\lambda$ being the spin-phonon coupling constant, and $\langle SS\rangle$ -- the spin-spin correlations.
Indeed, it is evident in the temperature dependencies of the parameters of the \Eg and \Ag phonons, as shown in \cref{fig:spin_phonon}(c,d) by green curves.
The strong change in the phonon line FWHM, especially pronounced for the \Eg mode, is due to coupling with magnetic excitons~\cite{mills_exciton_opticalphonon_1970}.

The temperature dependencies of the \Bone phonon mode parameters strongly differ from those for other phonons in the AFM phase as well (see \cref{fig:spin_phonon}(b)).
This is most apparent for the FWHM curve, which steadily decreases approaching \TN from above, at which point it starts to increase due to the spin-phonon effect as for other phonons.
However, at a particular temperature of around $T_*=23$~K it starts to drop rapidly again saturating below $T\approx15$~K at a very low value of just $1$~\cm.
The temperature dependence of the \Bone phonon frequency $\omega_\mathrm{B1g}$ also possesses features around \Tcr with a signature of an avoided crossing, as discussed below.
Noticeably, the \Bone line intensity in the ``forbidden'' configuration reaches its maximum at \Tcr as well [\cref{fig:polar_maps}(d)].
After careful examination of different excitations, we found that the temperature \Tcr is exactly the one at which the doubled frequency of the magnon crosses the frequency of the \Bone phonon, $2\omega_\mathrm{m}=\omega_\mathrm{B1g}$.

Since the FHWM of the phonon line in the Raman spectra is inversely proportional to the lifetime of the corresponding phonon mode, a pronounced increase in the lifetime below \Tcr, i.e., when $\omega_{B1g}<2\omega_m$, suggests that the latter condition blocks a particular phonon-mode dissipation channel.
Indeed, when $\omega_\mathrm{B1g} \geq 2\omega_\mathrm{m}$, it is possible for phonons to decay via the creation of two magnons with an energy of exactly half of their own and opposite wavevectors $k_m\geq0$, as illustrated in \cref{fig:magnetic_temp} using experimental data on the magnon dispersion in \CoF~\cite{martel_experimental_studies_1968}.
This mechanism is usually associated with phonon anharmonicity and is known as the cubic Klemens process~\cite{klemens_anharmonic_decay_1966,balkanski_Anharmonic_Effects_1983}.

Using the same approach, we analyze the temperature evolution of the parameters of the one-magnon mode summarized in~\cref{fig:magnetic_temp}.
Its frequency dependence follows the change of the magnetic moment with temperature and is well described by the Brillouin function \cref{fig:magnetic_temp}(a).
The FWHM dependence is nonlinear but monotonous.
No relaxation-related anomalies were seen in the FWHM, which can be rationalized by noting that it rises rapidly due to the thermal effects upon approaching \TN.
For example, the FWHM of this mode exceeds 10~\cm around \Tcr, which is five times higher that for the \Bone phonon at the same temperature.
However, the integral intensity of the magnon line possesses a well-pronounced maximum at \Tcr, see~\cref{fig:magnetic_temp}.
This is in contrast to the expected behavior of the magnon line in a Raman spectrum, yet it is in good agreement with the temperature dependence of the \Bone phonon line intensity in the ``forbidden" geometry [\cref{fig:polar_maps}(d)].

While the feature in the temperature dependence of the phonon FWHM at \Tcr is a manifestation of energy transfer from lattice to magnetic subsystem, the observation of the intensity of the \Bone phonon line in the ``forbiden'' configuration $b(cc)\overline{b}$ (\cref{fig:polar_maps}(d)) suggests that this transfer is mutual and contributes to the phonon creation process.
Indeed, in \cite{mashkovich_terahertz_light_2021}, it has been shown that coherently driven magnons enable the excitation of the phonon mode.
Our results allow further comprehension of this process.
First, our observations suggest that this process is intrinsic and requires neither strong resonant excitation of magnons nor coherence of the latter.
Second, by analyzing the azimuthal dependencies of the \Bone phonon line intensity shown in \cref{fig:polar_maps}(a) using the approach discussed in supplementary material, we obtain that the Raman tensor of the \Bone mode below \TN has one more component $\vert R_{cc}\vert\approx0.4c$ in addition to $R_{aa}=-R_{bb}=c$ allowed by the crystallographic symmetry.

In order to verify that the appearance of the \Bone intensity in the forbidden geometry described by the $R_{cc}$ tensor component agrees with the magnon mode symmetry, we analyze the direct product of the magnon representations.
Taking $E_g$ as the magnon representation (see~\cref{eq:raman_tensors:Eg}) for details, we obtain $E_g \otimes E_g = A_{1g} \oplus A_{2g} \oplus B_{1g} \oplus B_{2g}$.
Thus, it is feasible to get a response with a non-zero $cc$ component, namely $A_{1g}$, allowing non-zero intensity in $b(cc)\overline{b}$ polarization.

The lowering of the symmetry of the \Bone phonon Raman tensor below \TN, pronounced maxima in the intensity of both phonon and magnon lines, and anomalies in the frequency and FWHM of the phonon at \Tcr suggest that nonlinear coupling between lattice and magnetic excitation occurs under the condition $2\omega_\mathrm{m}\approx\omega_\mathrm{B1g}$.
The strength of this coupling could be analyzed qualitatively from the temperature dependence of the \Bone phonon frequency.
Indeed, close inspection of $\omega_\mathrm{B1g}$ in the vicinity of \Tcr reveals a signature of an avoided crossing~(\cref{fig:spin_phonon}).
This, in turn, allow us to characterize this type of nonlinear coupling as a strong one.
Note, however, that the type of coupling that we observe is different from the one-magnon-one-phonon one, where the energy of the magnon becomes comparable with that of the phonon and they form a mixed mode, resulting in an anticrossing~\cite{liu_direct_observation_2021} or the formation of the coupled magnetoelastic modes obtained through simultaneous diagonalization of the magnon-phonon Hamiltonian~\cite{white_diagonalization_antiferromagnetic_1965}.
The nonlinear interaction revealed in our study is closer to a two-magnons-one-phonon coupling, which was described by Dixon~\cite{dixon_lattice_thermal_1980}. 

The critical finding in our study is the intricate interaction of the magnon and \Bone phonon in the form of the two-magnons-one-phonon resonance in \CoF occurring at specific temperature \Tcr at which the frequency of the magnon becomes exactly twice as high as the frequency of the phonon.
Resonance interaction was observed both by anomalies in the frequency and FWHM of the \Bone phonon, integral intensity of the magnon, and modification of the \Bone phonon Raman tensor element in a form of non-zero intensity in the forbidden polarization.
It should be noted that this type of resonance is realized in a non-stimulated, non-coherent regime and being extremely strong.
We hope that our observations will shed new light on the complex dynamics in \CoF and in altermagnets in general and provide new avenues for control of both lattice and spin dynamics without spectral overlap of the individual excitations.
We expect that the application of a magnetic field, resulting in the a magnon energy change could be used for fine-tuning of the magnon-phonon resonance of this type.

The observed type of resonance resembles the mechanism of the resonant down-conversion of the Higgs to Goldstone phonon modes~\cite{juraschek_parametric_excitation_2020}, as opposed to the conventional parametric decay.
Also, the strong changes in decay rates of both phonons and magnons could be exploited for the Bose-Einstein condensation of magnons.

\begin{acknowledgments}
The technical assistance of the whole HFML technical team is greatly acknowledged, particularly P.~Alberts, F.~Janssen, and L.\,P.~Nelemans.
Fruitful discussions with E.\,A.~Mashkovich, J.~Mentink and Beatrice~T.~Crow, and support from E.\,A.~Arkhipova, D.\,A.~Andronikova, V.\,D.~Mamaeva-Niles, and O.\.M.~Palatina are acknowledged.
The support of the HFML-RU/NWO-I, member of the European Magnetic Field Laboratory (EMFL), is acknowledged.
M.\,A.\,P. acknowledges the support from Russian Science Foundation project 22-72-00039.

The authors declare that this work has been published as a result of peer-to-peer scientific collaboration between researchers. The provided affiliations represent the actual addresses of the authors in agreement with their digital identifier (ORCID) and cannot be considered as a formal collaboration between the aforementioned institutions.
\end{acknowledgments}

\newpage
\clearpage

{
\appendix

\section{\label{sec:appendix}Supplementary materials}

\subsection{Experiment}
\label{subsec:experiment}

High-quality single crystals originate from (TODO: add source)
The bulk crystals are pink in color, and rather transparent, thus an excitation mode-locked laser with wavelength of 660~nm (Thorus, LaserQuantum) was used to match the suggested transparency window to maximize the scattering volume, reduce laser overheating, and avoid any resonance effects in light scattering.
Careful examination of the laser power effect showed that the excitation powers up to 12~mW with a 50x objective leads to a negligible overheating while still maintaining a good signal-to-noise ratio.
For low-temperature azimuthal dependencies with a higher NA objective, the power was lowered down to 4 and 1~mW.
A set of Bragg filters (Optigrate) was used in beamsplitter/rejection filters backscattering configuration.
The scattered light was dispersed in a FHR1000 (HORIBA) 1~m spectrometer with 1200 l/mm grating, and registered by a liquid nitrogen cooled PyLoN CCD detector (Princeton instruments).

Temperature-dependent measurements were done using a helium flow cryostat (Oxford instruments) with temperature stability better than 1~K.
Samples were mounted on the cold finger with silver paste.
A plan achromatic long working distance objective MSPlan x50 (Olympus) was used to both guide the excitation light to the sample and to collect the scattered light.

Low-temperature azimuthal dependencies were both measured in a helium bath cryostat with samples in the exchange helium gas.
Singlet objective (NA = 0.68) with an effective focal length of 3.1~mm was used both for the excitation and collection of scattered light.
In all experiments, the same set of achromatic half-wave plates and Glan-Taylor polarizers (Thorlabs) were used to selectively probe specific elements of the Raman tensors.
To achieve a high level of signal-to-noise ratio and avoid overheating long exposure times were used for up to 1~h for some selected polarizations.
Experimental setups were controlled with the modular python framework qudi~\cite{binder_Qudi_Modular_2017}.

\begin{figure*}
    \centering
	\includegraphics[width=\textwidth]{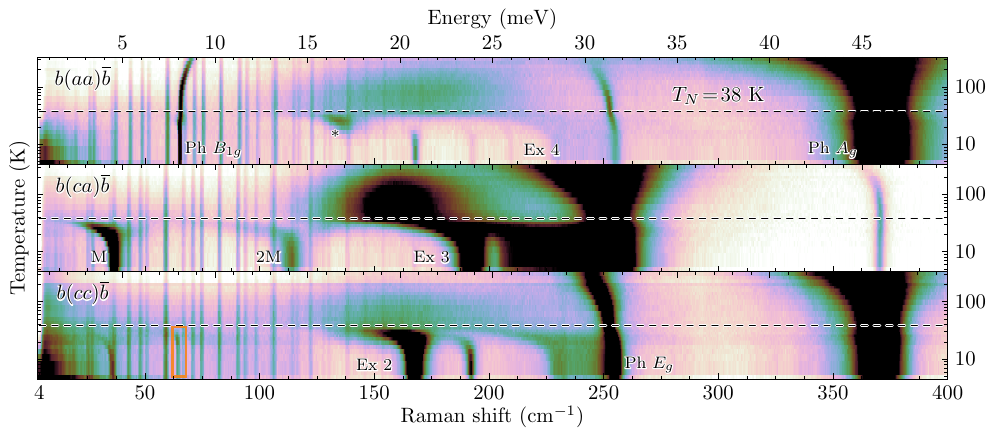}
	\caption{
	\label{fig:appendix_temperature_map} 
        Maps of the Raman scattering spectra vs. temperature.
        Most prominent excitaions are marked: M -- one magnon mode, 2M -- two magnon mode, Ex - magnetic excitons ($d$-$d$ transitions), Ph - phonons.
        Spectra are obtained in (a) $b(aa)\overline{b}$, (b) $b(ca)\overline{b}$, and (c) $b(cc)\overline{b}$.
        In $b(cc)\overline{b}$ spectrum, the nonzero intensity of the "forbidden" \Bone phonon line below \TN is highlighted by the orange rectangle. 
	}
\end{figure*}

\subsection{Raman tensor formalism \\and azimuthal dependencies of the Raman spectra
}
\label{subsec:raman_tensor}

Information about phonon frequencies is one of the basic ingredient for understanding both the static and dynamic dielectric properties of a crystal.
Moreover precise frequencies could be used as observables for \textit{ab initio} calculations, which, in its turn, could be used for further development of the models and a microscopic understanding. 

One of the important features of the lattice dynamics of \CoF is its highly entangled nature manifested as coupling of the phonon modes to the static and dynamic magnetism of this crystal.
This will be described in later subsections.
To obtain a full set of maximally useful information we performed angle-resolved measurements on two samples - $ab$- and $bc$-plane ones, which is a sufficient set for a uniaxial crystal.
Thus we are able to extract not only the usual parameters as frequency and FWHMs, but also the full Raman tensors elements themselves for each individual mode.
The exemplary results of such measurements -- polar dependencies of the Raman scattering intensity for the \Bone phonon line are shown in~\cref{fig:polar_maps}(a,b).  

For a known crystal structure it is possible to derive the Raman tensors, defining the selection rules for phonons~\cite{macfarlane_light_scattering_1970} and magnons~\cite{ariai_effects_linear_1982}.
These tensors for \CoF are listed below:

\begin{gather}
B_{1g} =
  \begin{pmatrix*}[r]
     c & 0 & 0 \\
     0 & -c & 0 \\
     0 & 0 & \phantom{-}0
  \end{pmatrix*}, \label{eq:raman_tensors:B1g}
\\
A_{1g} =
  \begin{pmatrix*}[l]
     |a|e^{i \phi_a} & 0 & 0 \\
     0 & |a|e^{i \phi_a} & 0 \\
     0 & 0 & |b|e^{i \phi_b}
  \end{pmatrix*}, \label{eq:raman_tensors:A1g}
\\
E_{g} =
  \begin{pmatrix*}[r]
    0 & 0 & 0 \\
    0 & 0 & e \\
    0 & e & 0
  \end{pmatrix*},
    \begin{pmatrix*}[r]
    0 & 0 & -e \\
    0 & 0 & 0 \\
    -e & \phantom{-}0 & 0
  \end{pmatrix*}, \label{eq:raman_tensors:Eg}
\\
\Gamma_3^+ + \Gamma_4^+ =
  \begin{pmatrix*}[r]
    0 & 0 & \delta \\
    0 & 0 & i\delta^* \\
    \epsilon & i\epsilon^* & 0
  \end{pmatrix*},
    \begin{pmatrix*}[r]
    0 & 0 & -\delta \\
    0 & 0 & i\delta^* \\
    -\epsilon & i\epsilon^* & 0
  \end{pmatrix*}, \label{eq:raman_tensors:magnon}
\end{gather}

Polar dependencies of the intensities were obtained via the following relation: 
\begin{gather}
	\label{eq:Raman intensity}
		I = |\bm{e}_i\,\mathcal{T}\,\mathcal{R}\,\mathcal{T}^{-1}\,\bm{e}_s|^2,
\end{gather}
where $\mathcal{R}$ is the Raman tensor, $\mathcal{T}$ is the rotation matrix, $\bm{e}_i$ and $\bm{e}_s$ are the polarization vectors of the excitation and scattered light, respectively. 

The values of the Raman tensor elements extracted from the polar dependencies (see, \cref{fig:polar_maps}) measured at 4.2~K are:
\begin{equation}
    \begin{gathered}
    \label{eq:element_values}
        a = 13.8, b = 11, c = 1.25,\\ e = 12.5, \delta = 2.55, \epsilon = 3.2
\end{gathered}
\end{equation}
The phase difference for the $A_{1g}$ mode is $\phi_a-\phi_b=\pi/2$.
As was suggested in~\cite{fleury_scattering_light_1968}, the Raman tensor for the magnon is not completely antisymmetric, due to a non-quenched orbital moment of the \ch{Co^2+} ions.

\subsection{Temperature dependencies of the phonons: anharmonic description and spin-phonon interaction}
\label{subsection:temperature_phonons}

It was known that rutiles are subjected to the spin-phonon interaction, and one of the earliest works was done on the fluorides \MnF and \FeF~\cite{lockwood_Spin_Phonon_1988}.
Later detailed work on \MnF showed that there is a pronounced SP effect on infrared-active phonons across the AFM transition as well as anomalous temperature-dependent behaviour of $E_{u1}$, which the authors attributed to incipient multiferroicty due to a lattice instability~\cite{schleck_elastic_magnetic_2010}.

However, using high resolution spectrometer we were able to track the spin-phonon effects with great precision.

To describe high temperature anharmonism we applied the well-known model with four-phonon relaxation processes~\cite{balkanski_Anharmonic_Effects_1983}:

\begin{gather}
	\label{eq:freq}
	\omega_i(T) = \omega_{i0} - A \left( 1 + \frac{2}{e^{\hbar \omega_{i0}/2k_BT} - 1} \right),
\end{gather}

\begin{gather}
	\label{eq:FWHM}
		\Gamma_i(T) = \Gamma_{i0} + C \left( 1 + \frac{2}{e^{\hbar \omega_{i0}/2k_BT} - 1} \right).
\end{gather}

The data extracted from our spectra in comparison to data from~\cite{cottam_Spinphonon_Interaction_2019} and results of the anharmonic curves according to~\cref{eq:freq,eq:FWHM} combined to Brillouin function describing magnetic contribution is shown in~\cref{fig:spin_phonon}.

We confirm an anomalous anharmonic temperature dependence of the \Bone mode at $T>$~\TN, namely its decrease with lowering temperature.
Such behaviour was associated with a large lattice instability~\cite{schleck_elastic_magnetic_2010}.
However, in the context of this paper we are more interested in the renormalization of the phonon modes across the paramagnet to antiferromagnet transition.

It is important to note that there are a few anomalies besides the usual spin-phonon effect.
First of all, the low frequency phonon experiences noticeable softening slightly below \TN with subsequent hardening at 30 K.
Such a non-monotonous effect is atypical for SP and it is possible to speculate about the contribution of other effects such as phonon-magnon coupling, associated, for example, with relaxation of the phonon in two magnons, which looks realistically considering the fact that the energy of this particular phonon is surprisingly close to double the energy of the magnon (at the gamma point).
A high degree of interconnection between \Bone phonon and magnon was also observed in~\cite{mashkovich_terahertz_light_2021}, supporting this hypothesis.
Actually we are able to directly show this intrinsic coupling by the observation of the temperature where these anomalies occur.
This temperature is \Tcr~$ = 23$~K and it is exactly when the energy condition $\omega_{B_{1g}} \geq 2*\omega_{mag}$ is satisfied.
However it is not clear what processes are exactly responsible for these anomalies, either relaxation of the phonon in two magnons (all with $\bm{k}=0$) or excitation of the phonon with two magnons.

To test the hypothesis of the mixed nature of the \Bone phonon excitation, we measured its integral intensity as a function of the temperature in the forbidden polarization $b(cc)\overline{b}$ as shown in~\cref{fig:B1g_phonon}.
As expected, according to its Raman tensor, it has a zero intensity in this particular polarization at elevated temperature.
Surprisingly, below the transition temperature it is pronounced, and its integral intensity also follows a non-monotonic dependency with a maximum around \Tcr, as shown in~\cref{fig:B1g_phonon}.
Since this feature shows a clear temperature dependence it does not seems to be a leak. 
It seems that the appearance of the signal for the forbidden polarization (it could be seen as lobes in the polar dependence in~\cref{fig:polar_maps}) is also associated with intrinsically coupled phonon-magnon dynamics. 

While frequency changes could be explained via changes in the force constants associated with the transition to the antiferromagnetic phase, the changes in the FWHM should reflect relaxation processes. 
Such anomaly, manifested in the \Bone mode is a non-harmonic FWHM temperature dependence, with an upturn just below \TN followed by a sharp decrease exactly at \Tcr.
The increase of the FWHM below \TN is observed on all phonons associated with the appearance of the additional relaxation paths such as single- or multi-magnon processes, low-energy magnetic exciton channels, etc.
However, at the crossing temperature the halfwidth noticeable decreases, indicating the increase of the phonon lifetime, which likely is associated with the energy condition where it is impossible for a single phonon to relax via creating two magnons of half its frequency.

To sum up, by carefully measuring the polar and temperature dependencies of lattice and spin modes, we were able to demonstrate that the dynamics of the lowest energy phonon with \Bone symmetry and magnon is intrinsically coupled even in non-resonant/non-coherent regieme in contrast to observations made by~\cite{mashkovich_terahertz_light_2021}.

\begin{figure}
    \centering
	\includegraphics[width=1.0\columnwidth]{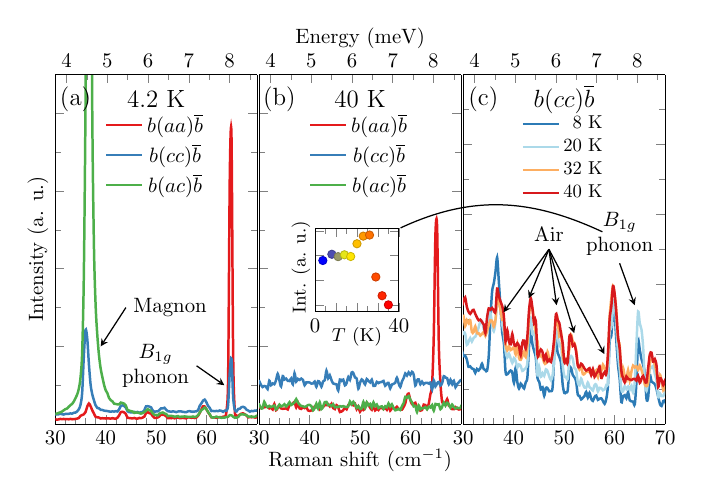}
	\caption{
	\label{fig:B1g_phonon} 
        Spectra of the magnon and $B_{1g}$ phonon at a) 4.2~K and b) 40 K at different polarizations.
        c) Temperature dependence of the $B_{1g}$ phonon in $b(cc)\overline{b}$ polarization.
        Its integral intensity in temperature range 4.2--35 K shown as inset in b).
        Note non-monotonous behaviour with maximum at $T_{*}$, see text for details.
	}
\end{figure}

\subsection{Full-widths of the magnon and \Bone}
\label{appendix:FWHM}
As mentioned previously we combined results from two experiments.
Temperature-dependent ones done with the use of the table-top flow cryostat at relatively high excitation powers and open entrance slit of spectrometer to minimize acquisition time.
The latter aspect leads to increased FWHM.
To get resolution limited (for a given combination of the spectrometer and diffraction grating) data we compared these measurements with data from bath cryostat in which polar dependencies were measured.
We were able to demonstrate that FWHM of the \Bone mode is extremely small, being about 0.6~\cm and, probably, still resolution limited.
It is possible to compare comparison to previous measurements reported in~\cite{cottam_Spinphonon_Interaction_2019}, where this value at lowest temperature around 10~K reported as 3.2~\cm.
By fitting digitized data with pseudovoigt lineshape as we done for our data the FWHM of 3~\cm is obtained.
Thus the overall resolution is about 5 times higher in comparison to most recent published Raman data, which is allowed us to track spin-phonon and spin-magnon interaction with great precision.
Moreover, such small value of the phonon halfwidth reflects its inherently long lifetime, probably reflecting the paucity of the decay channels.
It is also interesting to compare our data with recent THz time-domain measurements~\cite{mashkovich_terahertz_light_2021}, in which the width of the magnon line was reported to be $\approx 50$~GHz at 5~K.
For this the FWHM of the magnon line was measured in $b(ca)\overline{b}$ polarization at base temperature of 4.2~K, giving of 1.12~\cm$\approx 34$~GHz slightly lower but comparable with~\cite{mashkovich_terahertz_light_2021}, indicating that this FWHM could be treated close to an intrinsic characteristic of the magnon width.


\begin{figure*}
    \centering
	\includegraphics[width=0.8\textwidth]{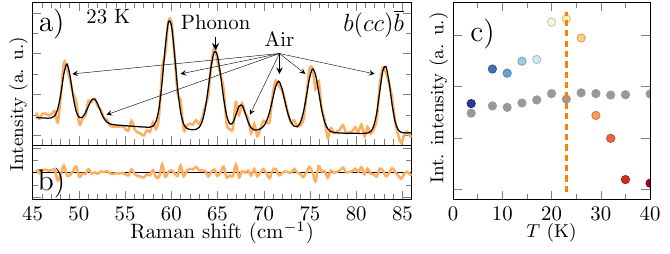}
	\caption{
	\label{fig:appendix_air_scattering} 
        a) Selected spectrum from the $b(cc)\overline{b}$ polarization at 23~K measured in flow cryostat during temperature-dependent run.
        Black curve is the model combining quadratic background and set of pseudo-voigt maxima.
        b) Residuals of the applied model and raw data.
        c) Comparison on the integral intensity of the \Bone phonon (colored circles) mode and averaged integral intensity of air lines (gray circles).
        This observation confirms intrinsic nature of the observed effect.
	}
\end{figure*}

\bibliography{CoF2_lattice}

\end{document}